\documentstyle[fleqn,run2col,epsfig]{article}

\def\beq{\begin{equation}}
\def\eeq{\end{equation}}  
\def\eq#1{{Eq.~(\ref{#1})}}
\newcommand{\as}{\alpha_S}

\newcommand{\AmS}{{\protect\the\textfont2
  A\kern-.1667em\lower.5ex\hbox{M}\kern-.125emS}}

\title{Higgs  and  Heavy  Quarks
Diffractive  Production
}

\author{Eugene  Levin \address{
 HEP Department, School of Physics,
 Tel Aviv University, Tel Aviv 69978, ISRAEL}
\address{
 Physics Department,
 Brookhaven National Laboratory,
 Upton, NY 11973 - 5000, USA
}%
  \thanks{Talk at ``RunII QCD and weak boson WS"  November  4 - 6,
Fermilab }        
 \thanks{Email: leving@post.tau.ac.il, elevin@quark.phy.bnl.gov}
}       
\begin{document}

\begin{abstract}
In this note we give  the highest of reasonable estimates for the value
of
cross section of the double Pomeron Higgs meson production  and suggest a 
new mechanism for heavy quark diffractive production which will dominate
at the Tevatron energies.

\end{abstract}

\maketitle

\begin{flushright}
BNL-NT-99/9 \\ 
TAUP-2615-99\\
\end{flushright}

\section{INTRODUCTION}
In this note we consider three  reactions
\begin{eqnarray}
p + p & \longrightarrow  & p + [LRG] + H + [LRG] + p \,\,;
\label{R1}\\
p + p & \longrightarrow  & X_1 + [LRG] + H + [LRG] + X_2 \,\,;
\label{R2}\\
p + p & \longrightarrow & b + \bar b + X + [LRG] + p
\,\,;\label{R3}
\end{eqnarray}
where LRG denotes the large rapidity gap between produced particles and
$X$ corresponds to a system of hadrons with masses much smaller than the
total energy.
The first two reactions are  so called double Pomeron production of Higgs
meson
while the third  is the single diffraction production of bottom -
antibottom pair. 

The goals of this note are the following:
\begin{enumerate}
\item\,\,\, To give the highest from reasonable estimates for the cross
sections of reactions \eq{R1} and \eq{R2}\,\,;
\item\,\,\, To summarize all uncertainties which we see in doing these
estimates\,\,;
\item\,\,\,  To show that there is a new mechanism of diffractive heavy
quark production ( \eq{R3} ) which is suppressed in DIS and dominates in
hadron-hadron
collision at the Tevatron\,\,;
\item\,\,\, To estimate the value of the cross section of reaction \eq{R3}
due to this new mechanism and to show that all attempts to compare the
diffraction dissociation in hadron-hadron collisions and DIS\cite{DDHDIS}
look unreliable without a detail experimental study of this process at
Fermilab.
\end{enumerate}
\section{DOUBLE POMERON HIGGS PRODUCTION}
\subsection{Inclusive Higgs production}
Inclusive Higgs production has been studied in many details
\cite{I1,I2,I3} for the Tevatron energies. The main source for Higgs is
gluon-gluon fusion which gives $\sigma (GG \rightarrow H) = 1 pb$ for
Higgs with mass $M_H =
10\, GeV$ \cite{I3}.  The reference point for our estimates is the cross
section of Higgs production due to W and Z fusion which is equal to 
$\sigma ( WW (ZZ) \rightarrow H) = 0.1 pb$ \cite{I3}. In  this process
we also expect the two LRG \cite{I2} and in some sense this is a
competing process for reactions of \eq{R1} and \eq{R2}.
\subsection{Double Pomeron Higgs production is a ``soft" process !!!}
\begin{figure}
\vspace{-0.4cm}
\begin{center}
\epsfig{file=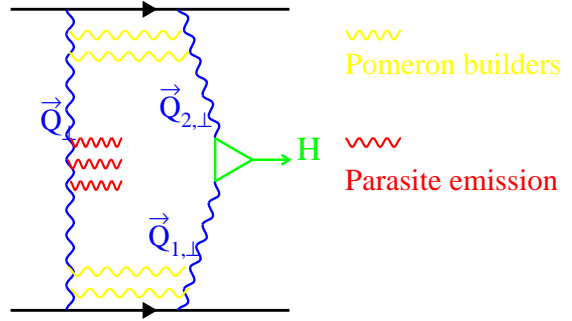,width=8cm}
\end{center}
\caption{\it Double Pomeron  Higgs production in QCD}
\end{figure}
Let us estimate the simplest digram for the DP Higgs production, namely,
Fig.1 without any of s-channel gluons. This diagram leads tpo the
amplitude
\beq \label{M1}
M (q q \rightarrow q H q) =
\eeq
$$
 \frac{2}{9} \,2\,g_H \int
\frac{d^2 Q_{\perp}}{Q^2_{\perp}\,Q^2_{1,\perp}\,Q^2_{2,\perp}}\,4
\alpha_S (Q^2_{\perp})\,( \vec{Q}_{1,\perp}\,\cdot\,\vec{Q}_{2,\perp} )\,.
$$
For reaction of \eq{R1}, $|t_1| = | \vec{Q}_{\perp} - \vec{Q}_{1,\perp}|
\,\approx\,|t_2| =     |
\vec{Q}_{\perp} -
\vec{Q}_{2,\perp} | \,\approx\,2/B_{el}$ and therefore, 
\beq \label{M2}
M (q + q \rightarrow q + H + q )\,\propto\,\int\frac{d^2
Q_{\perp}}{Q^4_{\perp}}.
\eeq
\eq{M2} has an infrared divergency that is regularized by the size of the
colliding hadrons. In other words, one can see that the simplest diagrams
shows that  DP Higgs production is a typical ``soft" process.
\subsection{The more the gluons the more the problems...}
In Fig.1 one can see that we have two sets of gluon which play a different
role. The first one is the gluons that connect $t$-channel lines. Their
contribution increases the value of cross section \cite{DP1,DP2,DP3,DP4}
\beq \label{G1}
\frac{d \sigma_P (pp \rightarrow pp H)}{d y}|_{y =0} \,=
\eeq
$$
\,\frac{4
\,g^2_H}{16^2 \,\pi^3}\,\,
\int d t_1 d t_2 g^2_{Pp} g^2_{Pp} \,e^{ \frac{B_{el}(s/M^2_H)}{2} (
t_1 + t_2 )} \left(\,\frac{s}{M^2_H}\,\right)^{2\,\Delta_P}
$$
\eq{G1} can be rewritten in the form
\beq \label{G2}
\frac{d \sigma_P (pp \rightarrow pp H)}{d y}|_{y =0}
= 
\eeq
$$
\frac{16}{\pi}\,\sigma(GG\,\rightarrow\,H)\,\left(
\frac{\sigma_{el}(s/M^2_H)}{\sigma_{tot}(s/M^2_H)} \right)^2 
$$
which is convenient for numeric estimates. However, first we need to find
the value of $\sigma(GG\,\rightarrow\,H) = g^2_H$. In inclusive production
the value of $ g^2_H$ has been calculated \cite{HX}
\beq \label{HX}
 g^2_H\,=\, \sqrt{2} G_F \alpha^2_S(M^2_H) N^2/9 \pi^2.
\eeq
However, I think that the scale of $\as$ for our process is not the mass
of Higgs but the ``soft" scale ( $\as(Q^2_0)$ with $Q^2_0 \approx 1
\,GeV^2$ ).  Indeed, using BLM procedure \cite{BLM}  we can include the
bubbles with large number of light quarks only in $t$-channel gluon line
which carry the ``soft'' transverse momenta. This gives a sizable effect
in numbers, since $\sigma(GG\,\rightarrow\,H)$ for $ M_H = 100\,GeV$ 
is equal to 1.16 pb ( $ \as(M^2_H)$ ) and to 20 pb ( $ \as(Q^2_0)$ ).
Taking the last value we have
\beq \label{N1}
\frac{d \sigma_P (pp \rightarrow pp H)}{d y}|_{y =0} =  2  pb
\eeq
This is our maximal value since all other effects related to gluon
emission suppressed the value of the cross section.
\subsection{ Cost of survival}
Actually, we have to multiply the cross section of \eq{N1} by two factors
to obtain the estimate for the experimental cross section
\beq \label{SP1}
{\frac d \sigma (pp \rightarrow pp H)}{d y}|_{y =0}
\eeq
$$ = S^2_{spect}
S^2_{par}
\frac{d \sigma_P (pp \rightarrow pp H)}{d y}|_{y =0}
$$
The first factor is the probability that there is no inelastic interaction
of the spectators in our process. I
The situation with calculation of this factor has been reported  in
this workshop \cite{LE} and the conclusion is that this factor
$S^2_{spect} = 0.07 \div 0.13 $ at the Tevatron energies. The discussion
for double Pomeron processes you can find in Ref. \cite{SP2}

The second factor in \eq{SP1} describe the probability that there is no
parasite emission in Fig.1 which leads to a process with 
hadrons in central rapidity region which do not come from the Higgs decay.
The generic formula for $S^2_{par}$ is
\beq \label{SP2}
S^2_{spect} = e^{- \,< N_G(\Delta y = ln(M^2_H/s_0)>}
\eeq
where $< N_G(\Delta y)>$ is the mean number of gluon in interval $\Delta
y$. In pQCD this number is large \cite{SP3} $ < N_G(\Delta y)
\,\approx\,8$ which leads to very small cross section for Higgs
production.  For ``soft'' double Pomeron production we can estimate the
value of $< N_G(\Delta y)$ assuming that the hadron production  is
two stage process: (i) production of mini jet with $p_t \approx 2 - 3
GeV$ and (ii) minijet decay in hadrons which can be taken from $e^+ e^-
\rightarrow  hadrons$ process. Finally, 
\beq \label{SP3}
 <N_G(\Delta y)> =\frac{ N_{hadrons}}{N( one\,\,\,\,\,
minijet)} \approx  2 \div 3 \,,
\eeq
which gives $ S^2_{parasite\,\,\,\, emission}  \approx 0.1$. 
\subsection{God loves the brave !!!}
Finally, we have
\beq \label{GOD1}
\frac{d \sigma (pp \rightarrow pp H)}{d y}|_{y =0} = 0.02  pb
\eeq
We can increase the  cross section, measuring reaction of \eq{R2}. Its
cross section is equal to
\beq \label{GOD2}
\frac{d \sigma (pp \rightarrow X_1 X_2  H)}{d y}|_{y =0} =
\eeq
$$
\frac{d \sigma
(pp \rightarrow pp H)}{d y}|_{y =0} \left(
\frac{\sigma^{SD}\cdot B_{el}(\sqrt{s}/M_H)}{4\,\sigma_{el}\cdot
B_{DD}(\sqrt{s}/M_H)} \right)^2 =
$$
$$
 3 - 4 \frac{d \sigma  
(pp \rightarrow pp H)}{d y}|_{y =0} = 0.06 \div 0.08 pb
$$
\subsection{Sensitive issues.}
\eq{GOD1} and \eq{GOD2} are our results. I firmly believe that they give
the maximum values of the cross sections which we could obtain from
reasonable estimates. However, I would like to summarize the most
sensitive points in our estimates:
\begin{enumerate}
\item\,\,\, The scale for running coupling QCD constant in cross section
of Higgs production. We  took the ``soft'' scale for our estimates.
However, it is a point which needs more discussion and even more it looks
in contradiction with our feeling, as I have realised during our last
meeting. My argument is the BLM procedure but
more discussions are needed;
\item\,\,\, We took $S^2_{spect} $ for double Pomeron processes the same
as for ``hard" LRG process. The justification for this is eikonal type
model \cite{SP2}, but it could be different opinions as well as direct
experimental data;
\item\,\,\, The estimates for $S^2_{par}$ is very approximate and we need
to work out better theory for this suppression.
\end{enumerate}
We would like also to mention that the new ideas on high energy
interaction such as the saturation of the gluon density at high
energy\cite{HDQCD},
will give a more optomistic estimates for the process of interest.

 \section{DIFFRACTIVE HEAVY QUARK PRODUCTION}  
The main observation is that there are two contributions for heavy quark
diffractive production (see \eq{R3}): (i) the first is so called
Ingelman-Schlein mechanism \cite{IS} which described by Fig.2-a and (ii)
the second one is closely related to coherent diffraction suggested in
Ref. \cite{CD} and which corresponds to Fig. 2-b. The estimates of both of
them have been discussed in Ref. \cite{DHQ1}. The main conclusion is that
the main contribution for the Tevatron energies stems from CD (see also
\cite{DHQ2,DHQ3} while the IS mechanism leads to the value of the cross
section in one order \cite{DHQ1,DHQ4}  less than CD one. on the other hand
in DIS the CD contribution belongs to the high twisdt and because of that
it
is rather small \cite{DHQ1,DHQ2}.

 \begin{figure}
\vspace{-0.4cm}
\begin{center}
\epsfig{file=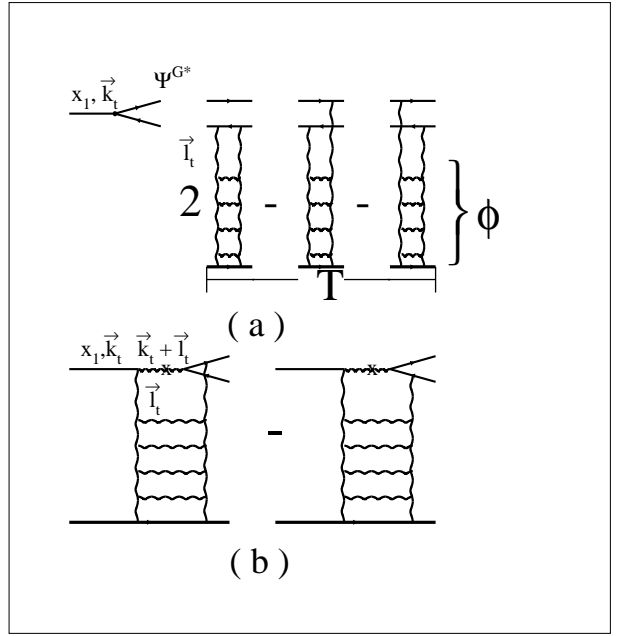,width=8cm}
\end{center}
\caption{\it The contributions for diffractive Higgs production: (a)
Ingelman-Schlein mechanism and (b) Coherent diffraction}
\end{figure}

Our conclusion is very simple. At the Tevatrom we has a good chance to
measure a new contribution to ``hard" diffraction which is small in DIS.
The typical values of the cross section is
$$
\frac{ d \sigma}{d Y}\,\,=\,\,\int^{\infty}_{p^{min}_t} d p^2_t
\int^{+\infty}_{-\infty}
d \Delta y \int^{\infty}_{0} d q^2_t \,\,\frac{d \sigma}{ d Y d \Delta y d
q^2_t
d p^2_t} 
$$
\beq \label{HQ1}
\approx 10^{-4} \div 10^{-10}\,\,\,\,\,\, for\,\,\,\,\,p_{t, min}
\,\,=\,\,5\,\div\,50\,\,GeV
\eeq
One can find all details in Ref. \cite{DHQ1}.

\section*{Acknowledgements:}
I am very grateful to A. Gotsman and U. Maor for encouraging optimism and
their permanent discussions on the subject.
My special thanks goes to Larry McLerran and his mob at the BNL for very
creative atmosphere and fruitful discussions.

This  research  was supported in part by the Israel Science
Foundation, founded by the Israeli Academy of Science and Humanities,
and BSF $\#$ 9800276. This manuscript has been authorized under
Contract No. DE-AC02-98CH10886 with the U.S. Department of Energy.

\end{document}